\documentclass[apl,graphicx, twocolumn]{revtex4-1}
\pdfoutput=1
\usepackage{amsmath}
\usepackage{graphicx}
\usepackage{natbib}
\usepackage{verbatim}
\usepackage[withbib, all]{authorindex}	
\usepackage[usenames]{color}
\usepackage{bm}
\usepackage{hyperref}
\hypersetup{
    colorlinks, linkcolor={blue},
    citecolor={blue}, urlcolor={blue},
}

\begin{document}

\title{Magnetic Resonance Force Detection using a Membrane Resonator}

\author{N.~Scozzaro}
\author{W.~Ruchotzke}
\author{A.~Belding}
\author{J.~Cardellino}
\author{E.~C.~Blomberg}
\author{B.~A.~McCullian}
\author{V.P.~Bhallamudi}
\author{D. V.~Pelekhov}
\author{P. C. ~Hammel}\email[]{Hammel@physics.osu.edu}\affiliation{Department of Physics, The Ohio State University, Columbus, Ohio 43210, USA}

\date{\today}

\begin{abstract}
The availability of compact, low-cost magnetic resonance imaging instruments would further broaden the substantial impact of this technology.  We report highly sensitive detection of magnetic resonance using low-stress silicon nitride (SiN$_x$) membranes. We use these membranes as low-loss, high-frequency mechanical oscillators and find they are able to mechanically detect spin-dependent forces with high sensitivity enabling ultrasensitive magnetic resonance detection.  The high force detection sensitivity stems from their high mechanical quality factor $Q\sim10^6$ \cite{Chakram2014, Zwickl2008} combined with the low mass of the resonator. We use this excellent mechanical force sensitivity to detect the electron spin magnetic resonance using a SiN$_x$ membrane as a force detector. The demonstrated force sensitivity at 300 K is 4 fN/$\sqrt{\mathrm{Hz}}$, indicating a potential low temperature (4 K) sensitivity of 25 aN/$\sqrt{\mathrm{Hz}}$. Given their sensitivity, robust construction, large surface area and low cost, SiN$_x$ membranes can potentially serve as the central component of a compact room-temperature ESR and NMR instrument that has superior spatial resolution to conventional approaches.
\end{abstract}


\maketitle 

\section{Introduction}
Magnetic resonance is a powerful tool that has revolutionized the fields of medicine, chemistry, and physics. Modern nuclear magnetic resonance (NMR) and magnetic resonance imaging (MRI) apparatuses utilize technology that have benefited from six decades of development, but still they have their limitations: the cost and size of NMR apparatuses are barriers that limit accessibility to the technology, and the spatial resolution of MRI is at best a few microns \cite{Ciobanu2002,Lee2001,Glover2002,Moore2015}. A design that is less expensive, more compact, and has superior resolution to conventional NMR/MRI would be a significant achievement that could enable a wide variety of new applications.

Magnetic resonance force microscopy (MRFM) is a technique based on mechanical detection of magnetic resonance signals. It has demonstrated imaging resolution far beyond that of conventional inductive-based MRI, achieving nuclear-spin resolution better than ten nanometers \cite{Degen2009,Nichol2012}. The central component of MRFM is a mechanical resonator with high force sensitivity. The resonator is used to measure the force of interaction between the sample containing electron or nuclear spins, and a probe magnet with a strong field gradient.  Depending on the experimental configuration, either the sample or the probe magnet is placed directly on the resonator \cite{Degen2009,Rugar2004} while the other component is in a fixed position in close proximity. The force of the probe-sample interaction is measured by detecting the displacement of the resonator via optical interferometry.  Thermal force noise as low as  0.82 aN/$\sqrt{\mathrm{Hz}}$ was achieved \cite{Mamin2001} using MRFM, culminating in an experiment that demonstrated single electron spin detection \cite{Rugar2004}. At present such a high sensitivity has been achieved by using ultrasoft cantilevers with spring constants $k$ as low as 110 $\mu$N/m  \cite{Rugar2004}. While such cantilevers deliver exceptional force sensitivity, their use presents challenges: they are very fragile, sample preparation in the sample-on-cantilever geometry is difficult due to its small size, and such cantilevers are not commercially available.

Here we demonstrate that SiN$_x$ membranes present a viable alternative to ultrasoft cantilevers as a sensitive force detectors for MRFM applications.  Such membranes exhibit a number of attractive properties including high sensitivity, robust mechanical properties, commercial availability, low cost, and a large surface area. The force noise of membranes can be as low as 8 aN/$\sqrt{\mathrm{Hz}}$ \cite{Zwickl2008}, which is within an order of magnitude of the highest sensitivity demonstrated so far for an ultrasoft cantilever \cite{Mamin2001}. While less sensitive, the membrane is much less fragile because it is surrounded on all sides, unlike a cantilever supported only at one end. As a result, unlike the ultrasoft cantilever, the membrane is much less susceptible to bending and twisting which otherwise can be detrimental for interferometric displacement detection. Samples can be quickly prepared on membranes utilizing similar sample preparation techniques as transmission electron microscopy (TEM), including application to the membrane by micropipette, or immersion in fluid on a glass slide. Finally, with their large surface area, membranes can accommodate a wider range of sample sizes and provide a larger target for interferometry than ultrasoft cantilevers.

One further advantage of membranes is their high natural frequency. The fundamental frequency of membranes can be in the MHz range, which enables resolving faster spin dynamics, and opens the door to new experiments. For example, since the Larmor frequency of nuclear spins in low field is also in the MHz range, there is the possibility of matching the membrane's mechanical resonance with the spins’ nuclear magnetic resonance frequency. This matching could overcome the current restriction to measuring only the $z$-component of the magnetization and allow force detection of its transverse component. This enticing possibility would provide MRFM access to the powerful array of imaging tools developed for pulsed NMR.  Transverse detection of Larmor precession, known as direct-detection or ``spin precession imaging'' \cite{Sidles1995,Sidles1992}, could furthermore be accomplished without the necessity of an RF-generator. As the membrane oscillates, the spin sample on the membrane is physically displaced in the presence of a strong field gradient, which naturally generates the large oscillating magnetic field needed to excite the magnetic resonance signal. This innovation would aid in simplifying the MRFM apparatus. The apparatus could thus be reduced to three main components: a membrane, optical fiber-based displacement detection, and a magnetic particle on a translation stage. 

\section{Experiment}
\subsection{Experimental details and setup}

The MRFM experiment is performed by creating an oscillating force on a sample placed in the center of the membrane. The oscillating force is generated by modulating the sample magnetization using magnetic resonance, in the presence of a field gradient. The force drives the membrane at its natural frequency to an amplitude $A=FQ/k$, where $F$ is the force, $k$ is the spring constant, and $Q$ is the quality factor of the membrane. We use the cyclic-saturation resonance protocol \cite{Rugar1992,Wago1997} to measure a small particle of diphenyl picrahydrazyl (DPPH) on a membrane. DPPH is a well-known organic molecule that exhibits a strong electron paramagnetic resonance (EPR) signal.

\begin{figure}
\includegraphics[width=1\linewidth]{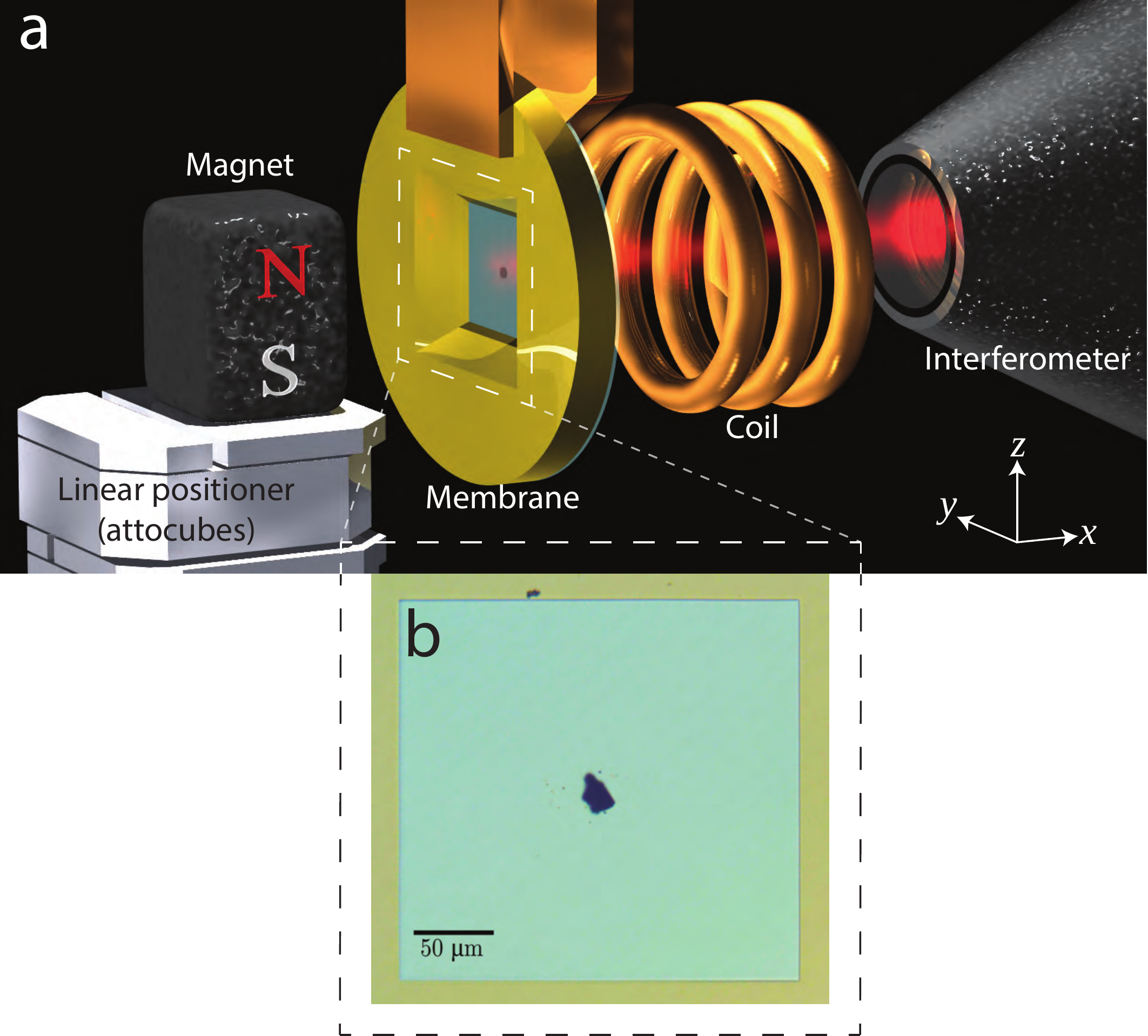}
\caption{\textbf{Experimental schematic}: (a) Schematic of the experimental setup. From the left, the interferometer laser passes through the copper coil and is incident on the membrane. A 20 $\mu$m particle of DPPH is placed in the center of the membrane. A permanent magnet produces both the polarizing field $B_0$  and a magnetic field gradient, which generates a force on the electron spins in the DPPH.  (b) Photomicrograph of the 30 nm thick, 250 $\mu$m side-length silicon nitride membrane (inner turquoise square), surrounded by the silicon support. In the center of the membrane is the piece of DPPH is attached using a small amount of G1 epoxy. The bare membrane exhibits a frequency of 1.35 MHz, and the loaded membrane exhibits a frequency of 644 kHz.  }
	\label{fig:schematic}
\end{figure}

The magnetic resonance signal is detected by measuring the displacement of a SiN$_x$ membrane by means of a fiber optic interferometer, aligned as shown in figure \ref{fig:schematic}. The interferometer uses 1550 nm laser light and is focused down to a 10 $\mu$m  spot adjacent to the DPPH particle. The optical power incident on the membrane was about 80 $\mu$W. A 2.5 turn, 350 $\mu$m diameter copper resonance coil generates $B_1$, the the RF field.  The coil is centered on the membrane and is stub-tuned to a frequency $\omega_0/2\pi= 3.010$  GHz , setting the magnetic resonance condition of $\omega_0/\gamma=1070$ G, where $\gamma$ is the electron gyromagnetic ratio. With an input power of 100 mW, the coil produces $B_1=4.7$ G. On the opposite side of the membrane, a two-axis piezoelectric stage (attocubes  \cite{attocube})  positions a rectangular NeFeB magnet, which provides both the polarizing magnetic field ($B_0$) and a field gradient of $G=0.2$ G/$\mu$m. The instrument operates at a pressure of $10^{-6}$ torr. 

 Using the coordinate system in figure \ref{fig:schematic}, the force on the electrons in the DPPH is given by $F=\mu_z \frac{\mathrm{d}B_z}{\mathrm{d}x}$, where $\mu_z=M_z V$, and $V$ is the volume of the DPPH particle.  The Bloch equations give the magnetization 
\begin{equation}
	M_z=M_0 \left(1-\frac{\gamma^2 B_1^2 \tau^2}{1+\left(\gamma B_0-\omega \right)^2 \tau^2+\gamma^2 B_1^2 \tau^2} \right),
\label{equation:Bloch equation}
\end{equation}
where $M_0=\frac {\chi_0 B_0}{\mu_0}$  is the thermal magnetization, $B_0$ is the magnetic field in the $z$-direction generated by the permanent magnet, $\chi_0=2.5\times 10^{-5}$  is the susceptibility of DPPH, $\mu_0$ is the permeability of vacuum, $\tau=62$ ns is the spin relaxation time, and $\frac{\gamma}{2\pi}=28$ GHz/T is the electron gyromagnetic ratio.  To create the oscillating force, we induce an oscillating moment at the membrane frequency by modulating the frequency $\omega$ of $B_1$ such that $\omega(t)=\omega_0+\Omega\sin{(2\pi f_ct)}$. This results in a time-varying magnetization $M_z (t)$ whose Fourier component $M_1$ at the membrane frequency given by $M_1=\Omega \frac{\partial M_z}{\partial \omega}$ \cite{Rugar1992}. The derivative leads to a bipolar line shape of the force as a function of $B_0$. 

The magnetic resonance signal is measured by varying the position of the permanent magnet and hence the magnitude of the applied magnetic field experienced by the sample on the membrane. The region of the magnet's field where the resonance condition is satisfied ($B=1070$ G) is referred to as the ``resonant slice,'' the thickness of which is $\Delta B/G=50$  $\mu$m, where $\Delta B=10$ G is the linewidth of DPPH. Due to the gradient, each spin in the resonant slice experiences a slightly different field, so the force as a function of magnet position is an integral as described in reference \cite{Wago1997}; see the appendix.

\subsection{Force noise}
The central component of the MRFM apparatus is a sensitive mechanical oscillator which is employed as a force detector whose force sensitivity is limited by thermal force noise $S_F$ given by
\begin{equation}
	S_F^{1/2}=\left (\frac{2kk_B T}{\pi Q f_0}\right )^{1/2},
\label{equation:thermal_force_noise}
\end{equation}
where $ k$ is the spring constant, $ k_b$ is the Boltzmann constant, $T$ is the temperature, $Q$ is the quality factor, and $f_0$ is the natural frequency. It is illuminating to cast equation \ref{equation:thermal_force_noise} in terms of intrinsic membrane parameters such as thickness, side length, tensile stress, and density; $t$, $L$, $\sigma$, and $\rho$, respectively. The frequency is given by $f_0=\sqrt{\frac{\sigma}{2\rho L^2 }}$, and the spring constant is given by $k=\frac{\pi^2 \sigma t}{2}$ \cite{Zwickl2011}, yielding
		
\begin{equation}
\begin{split}
S_F^{1/2}=\left (2 \rho \sigma \right )^{1/4} \left ( \frac{ \pi L tk_B T}{Q} \right )^{1/2} \\ 
\sim \left (\rho \sigma \right)^{1/4} \left (\frac{ t^3 k_B T}{L}\right )^{1/2},
\label{equation:fundamental_thermal_force_noise}
\end{split}
\end{equation}
where in the last step we further simplified the expression using the experimentally observed relation \cite{Chakram2014} that quality factor goes as $Q\sim(L/t)^2$ for $L/t<10^5$. Equation (\ref{equation:fundamental_thermal_force_noise}) shows that to minimize the thermal force noise, it is desirable to have a low stress, low density material, large side length, and most critically, a very thin membrane. 

To this end, we use a 30 nm thick, $0.25$ mm side length membrane for our experiments ($L/t=8\times 10^3$). The membranes we used are made of low-stress SiN$_x$ from the manufacturer Structure Probe Inc. (SPI) \cite{SPI}, and exhibit a natural frequency of 1.35 MHz. Given the density of low-stress silicon nitride, $\rho = 3.1$ g/cm$^3$, this corresponds to a stress of $\sigma=654$ MPa, and a spring constant of $k=97$ N/m. After attaching a DPPH particle to the center of the membrane with epoxy, the frequency of the membrane decreased to roughly 644 kHz. This corresponds to an increase in mass of $4.65\times 10^{-12}$ kg, or a particle diameter of $19$ $\mu$m, consistent with the optical image of the membrane particle in figure \ref{fig:schematic}b.

\subsubsection{Membrane characterization}
\begin{figure*}
\includegraphics[width=.8\textwidth]{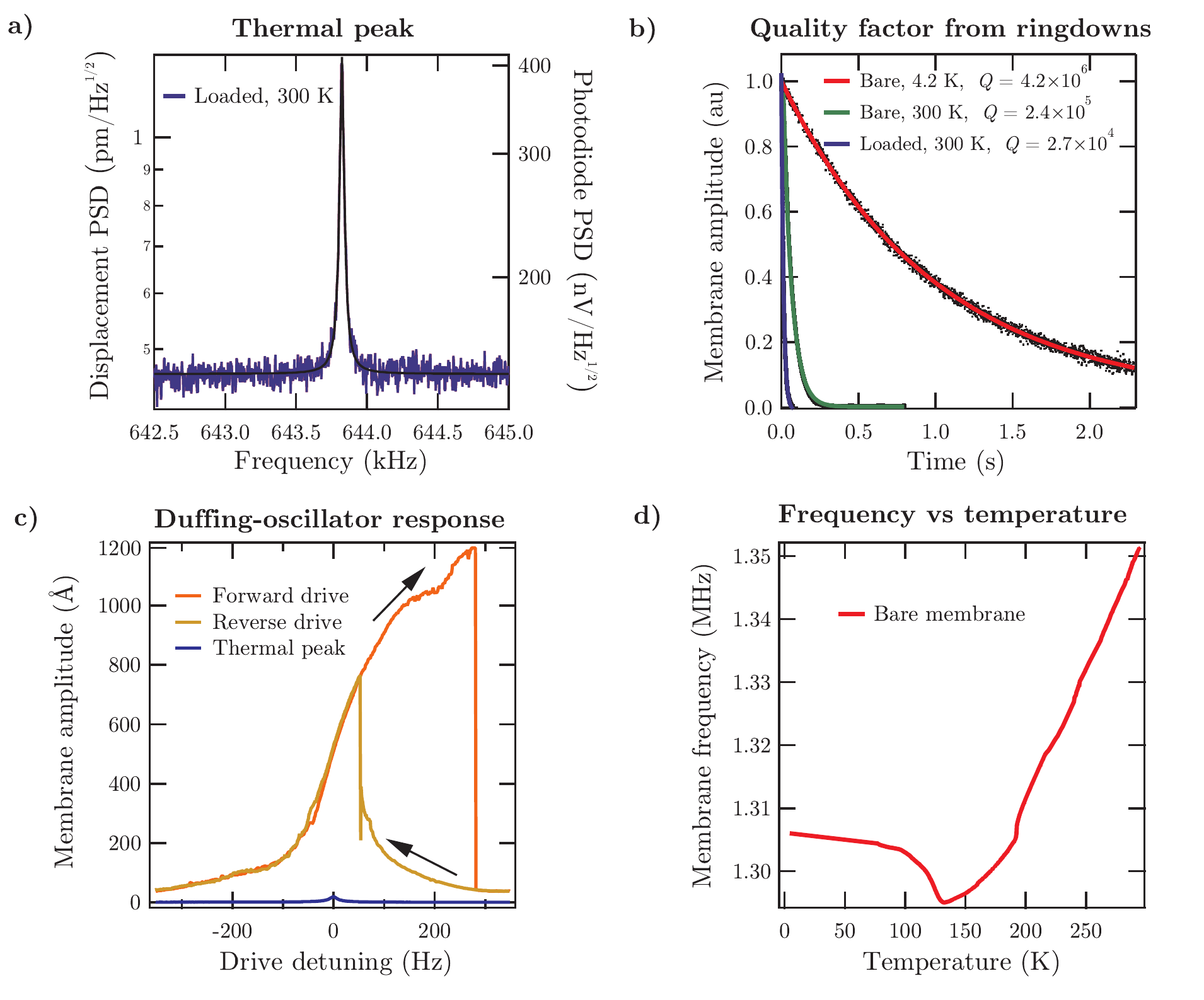}
\caption{\textbf{Characterization of the membrane signal} (a) Thermal peak of the DPPH-loaded membrane measured at 300 K. The membrane frequency dropped from 1.35 MHz when bare (as in panel (d)) to $\sim644$ kHz when loaded. The peak is fit to equation (\ref{equation:thermal_peak}) by fixing $T=300$ K, and $Q=27,000$ from the ringdown measurement in panel (b). The free parameters are the spring constant, the center frequency, and the white background interferometer noise, yielding $k=75$ N/m, $f_0=643.8$ kHz, $S_{bgd}^{1/2}=0.46$ pm/$\sqrt{\mathrm{Hz}}$. (b) Membrane ringdown measurements under 3 conditions: (1) Bare membrane at 4 K, (2) Bare membrane at 300 K, and (3) DPPH-loaded membrane at 300 K. We find that cryogenic temperatures increase the quality factor of the membrane by over an order of magnitude, from $2.4\times10^5$ to $4.2\times10^6$. On the other hand, loading the membrane with a particle of DPPH drastically reduces the quality factor down to $2.7\times10^4$. (c) Driving the membrane to large amplitudes shows the well-known Duffing-oscillator response, which results from a cubic term in addition to the linear term in the spring constant. The non-linearity results in an increase in resonance frequency for large drive amplitudes, as well as causing the sharp discontinuities, and the hysteresis in scanning forward or reverse. (d) The membrane’s frequency as a function of temperature. This non-monotonic behavior is likely related to the differential thermal expansion coefficient  between the silicon nitride thin film and the silicon substrate \cite{Martyniuk2006,Swenson1983}. }
	\label{fig:characterizationPlots}
\end{figure*}

We measure the displacement of the membrane using a fiber-optic interferometer and we fit the thermal peak and background (figure \ref{fig:characterizationPlots}a) using the equation
\begin{equation}
S_x^{1/2}=\left( \frac{f_0 k_B T}{2 \pi k Q \left( \left ( f-f_0 \right )^2+ \left( \frac{f_0}{2Q} \right)^2  \right)}+ S_{\mathrm{bgd}}\right)^{1/2}
\label{equation:thermal_peak}	
\end{equation}
where $S_{bgd}$ is the  displacement noise floor of our interferometer in units of m$^2$/Hz. Fixing $T=300$ K and $Q=27,000$ from a ringdown measurement (figure \ref{fig:characterizationPlots}b), the fit yields $f_0=643,800$ Hz, $S_{bgd}^{1/2} =0.46$ pm$/\sqrt{\mathrm{Hz}}$, and $k=75$ N/m, in reasonable agreement with the spring constant of 97 N/m predicted based on the loaded frequency of the membrane. From the equipartition theorem the membrane's thermal RMS amplitude is $x_{rms}=\sqrt{\frac {k_B T}{k}}=7.4$ pm.

We perform ringdown measurements (figure \ref{fig:characterizationPlots}b) to extract the membrane's quality factor by driving the membrane, ceasing the drive, then measuring the amplitude decay with a lock-in amplifier. We fit the amplitude decay to an exponential, $Ae^{-\frac{t}{\tau}}$, and use $Q=\pi f_0 \tau$ to extract $Q$. The bare membrane exhibited a quality factor of $2.4\times10^5$ at room temperature, which increased to $4.2\times 10^6$ upon cooling the system to low temperature. The impressively large $Q$ at low temperature (4 K)  is in agreement with other measurements \cite{Chakram2014, Zwickl2008}. Loading the membrane with the DPPH particle significantly reduced the quality factor to $Q=2.7\times 10^4$. Using these quality factors, equation (\ref{equation:thermal_force_noise}) yields a thermal force noise of 4 fN/$\sqrt{\mathrm{Hz}}$ at room temperature, and 25 aN/$\sqrt{\mathrm{Hz}}$  for the bare membrane at low temperature. While we have found that adding a large particle will drastically reduce the quality factor of the membrane, thereby making the membrane less sensitive, we have also placed particles on the membrane (~1 $\mu$m) small enough that they did not affect the quality factor. 

Using a piezoelectric element we can sinusoidally drive the membrane to large amplitudes as shown in figure \ref{fig:characterizationPlots}c. Due to a cubic term in the spring response of the membrane, $F=kx+\beta x^3$, the equation of motion is non-linear and is given by  $\ddot{x}+\left( \frac{\omega_0}{Q}\right) \dot{x}+\omega_0^2 +\beta x^3 = F_0 \sin{\omega t}$ \cite{Landau}, where $\beta$ is a parameter that describes the strength of the nonlinearity. This results in the well-known Duffing-oscillator response that is a deviation from the Lorentzian response of a linear oscillator. The shape of the signal can be understood in that the membrane effectively becomes stiffer for large amplitudes of oscillation, which causes an increase in the natural frequency of the membrane. The non-linearity also causes hysteresis such that driving from below resonance to above resonance yields a different response than sweeping from above resonance to below resonance. While it is possible to apply feedback to cancel the Duffing non-linearity \cite{Nichol}, we instead work at sufficiently low amplitudes that our experiment is performed in the linear regime. 

As the bare membrane is cooled to cryogenic temperatures (figure \ref{fig:characterizationPlots}d), we observe that the frequency decreases for temperatures between 300 K and 130 K, then increases between 130 K and 4 K. We attribute this non-monotonic behavior to the differential coefficient of thermal expansion between the SiN$_x$ thin films and the silicon substrate \cite{Martyniuk2006,Swenson1983}. For high temperature SiN$_x$ thin film deposition on silicon, the coefficient of thermal expansion for silicon is greater than that of SiN$_x$ which results in a decrease in the stress of the membrane as it cools, and correspondingly the frequency decreases. As the temperature continues to decrease past 130 K the coefficients cross, reversing the effect: the stress increases, and membrane frequency increases. 

\section{Results}

\begin{figure}
\includegraphics[width=1\linewidth]{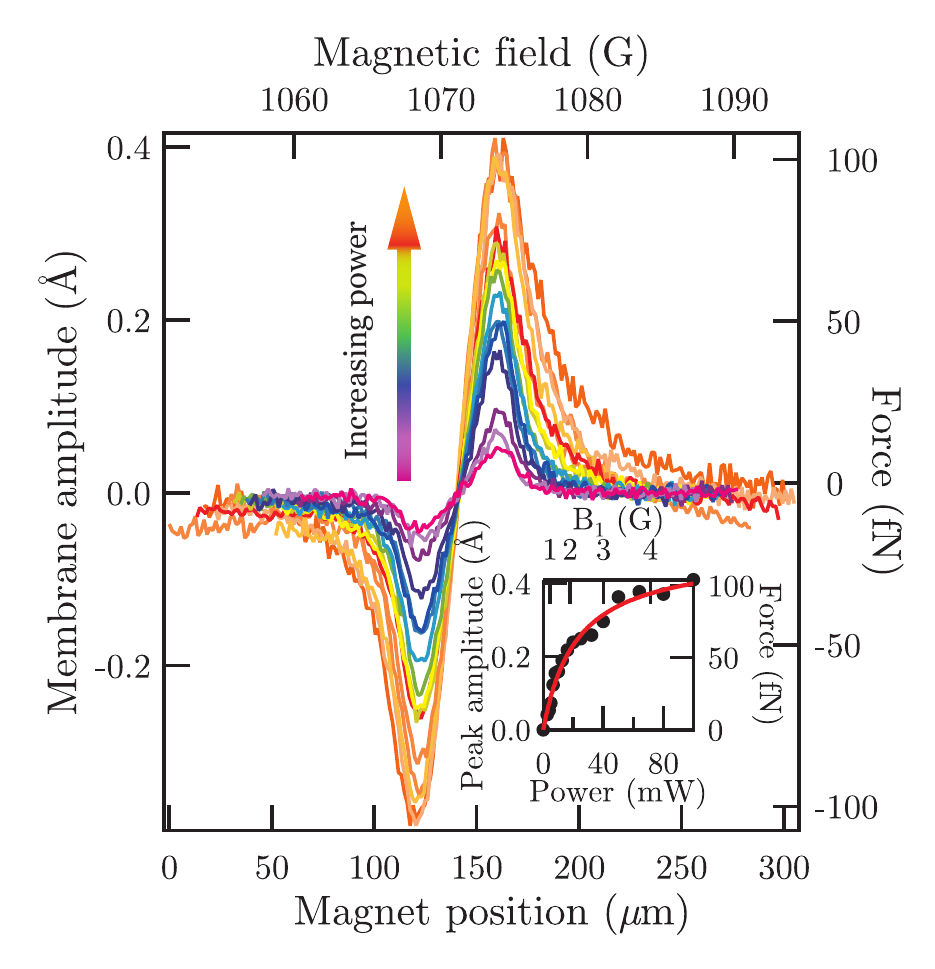}
\caption{\textbf{Cyclic saturation power dependence} Microwave power is increased from 2.5 mW to a maximum of 100 mW, corresponding to a $B_1$ of 4.7 G. The signal grows with increasing microwave power, peaking with an amplitude of 0.4 $\text{\AA}$ corresponding to a force of 100 fN. Each curve is fit with equation \ref{equation:cyc_sat_fit_eqn}, and the peak force is extracted and plot in the inset. The overlaid curve in the inset is from the maximum value of equation \ref{equation:cyc_sat_fit_eqn} as a function of microwave power.  }
\label{fig:Power_dependence}
\end{figure}

Figure \ref{fig:Power_dependence} shows the measured DPPH magnetic resonance signal as a function of microwave power. The experimental parameters are $\omega_0/2\pi=3.010$ GHz, $\Omega/2\pi=10$ MHz, $T=300$ K. We take multiple measurements at each point, and use more averaging for smaller powers. We fit each curve with equation (\ref{equation:cyc_sat_fit_eqn}) (appendix) and extract the peak forces, which are shown in the inset. The magnetic force drives the membrane to a maximum vibration amplitude of 0.4 angstroms as shown on the right axis for a power of 100 mW, corresponding to $B_1=4.7$ G. The gradient we extract is approximately $G=0.2$ G/$\mu$m, and the force corresponds to a polarized moment of $\mu_z = 10^{-14}$ J/T, or $10^9$ electron spins. 

\begin{figure}
\includegraphics[width=1\linewidth]{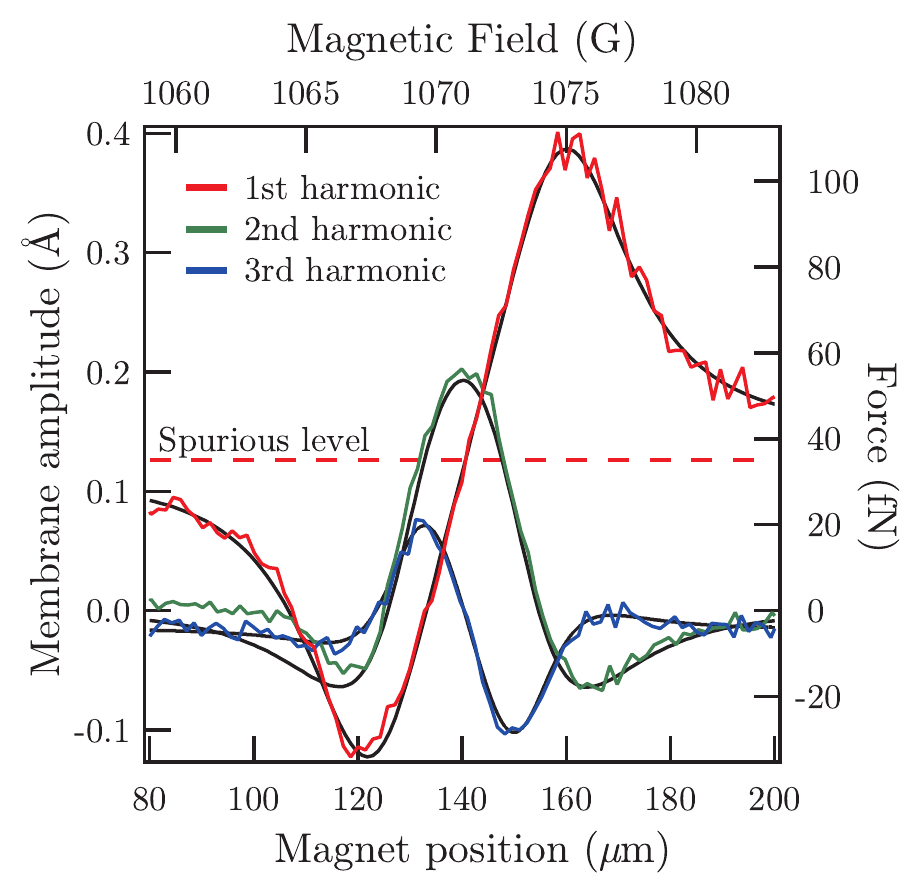}
\caption{\textbf{Higher harmonic data}
Measurement of the first three harmonic spin signals with a power of 32 mW. The first harmonic data (red) is identical to that shown in figure \ref{fig:Power_dependence}, and is provided for reference.The second harmonic and third harmonic spin signals are the green and blue curves respectively. Second and third harmonic measurements are performed by modulating the microwave frequency at half and one third the membrane frequency, respectively, while monitoring the membrane amplitude at the membrane’s natural frequency. The first, second, and third harmonic signals are proportional to the first, second, and third derivatives of the magnetization, respectively.  
  }
\label{fig:Second_harmonic}
\end{figure}

We also measure higher harmonic spin signals (figure \ref{fig:Second_harmonic}) which have been shown to reduce spurious noise \cite{Rugar1992}, and provide additional confirmation of the origin of the signal. Due to the nonlinearity of equation \ref{equation:Bloch equation} with respect to RF modulation, magnetic resonance signals are generated not just at the modulation frequency $\omega_m$, but also at integer multiples: $2\omega_m$, $3\omega_m$, etc. We measure these higher harmonic signals by setting the microwave modulation rate at half the membrane’s frequency $\omega_m= (2\pi f_c)/2$, or one-third, etc., resulting in a signal at the membrane’s frequency which we detect. The first three harmonic signals are given by  $M_1=\Omega \frac{\partial M_z}{\partial \omega}$, $M_2=\frac{\Omega^2}{4} \frac{\partial^2 M_z}{\partial\omega^2}$, $M_3=\frac{\Omega^3}{24}  \frac{\partial^3 M_z}{\partial \omega^3}$; each additional harmonic is one higher derivative.  For modulation at the membrane frequency and 32 mW of microwave power, spurious coupling appears as a constant oscillation amplitude offset of 0.1 $\text{\AA}$ in our in-phase lock-in channel, on top of which rides the 0.3 $\text{\AA}$ spin signal. We subtract this background from the data in figure \ref{fig:Power_dependence}. In contrast, spurious coupling is nearly eliminated using modulation away from the membrane’s frequency, and no background subtraction is necessary in figure \ref{fig:Second_harmonic}.

\section{Discussion}
We now address the potential for detection of nuclear spins with membranes. Detection of nuclear moments is more difficult than detecting electrons for two reasons: (1) The moment of nuclear spins is much smaller than that of electrons, and (2) a very large (> 100 G) $B_1$ is needed to adiabatically invert nuclear spins at the membrane frequency. 

The minimum number of spins $n$ that can be measured using cyclic saturation is set by Curie's law and the thermal force noise of the membrane \cite{Rugar1994}, $n=\frac{\sqrt{2}S_F \Delta \nu k_B T}{\mu_N^2 G}$, where $\mu$ is the nuclear or electron spin moment, and $\Delta\nu$ is the detection bandwidth. Given that the electron has a moment 660 times larger than that of a nuclear spin such as a proton, it is readily seen that a significantly smaller number of electron spins are required to achieve the same signal as nuclear spins. For our membrane under the conditions presented in this paper, the minimum number of total protons we would be able to detect at room temperature is  $5.5\times 10^{16}$ protons, corresponding to a voxel size of 475 $\mu$m (assuming a spin density of $5.1\times 10^{28}$ protons$/\mathrm{m}^3$ protons/m$^3$). However, with experimentally accessible parameters such as a gradient of 40 G/nm \cite{Degen2009}, a polarizing field of 5 T, and operating at low temperature, the membrane can potentially be used to detect 3.7$\times 10^{4}$ spins corresponding to a voxel size of 40 nm. Nuclear spin resolution as high as 10 nm has been experimentally demonstrated \cite{Degen2009} with cantilevers that exhibit an order of magnitude better sensitivity than membranes. 

The second challenge is that a very large $B_1$ is required. When the relaxation time of spins is longer than the period of the mechanical oscillator, cyclic saturation cannot be used to generate an oscillating force for MRFM detection. Instead, adiabatic inversion of spins is necessary, which requires $B_1\gg \sqrt{\frac{4f_c \Omega}{\gamma^2}}$, where $\Omega$ is the depth of frequency modulation. The MRFM signal grows linearly with $\Omega$, so it cannot be made arbitrarily small. Since the frequency of the membrane is very large, this means that a very large $B_1$ is required, over 100 G. Nichol et. al  used a 240 nm wide, 100 nm thick constriction in a silver wire to achieve a $B_1$ of 88 G for MRFM detection \cite{Nichol2012,Nichol}; however, achieving such high $B_1$ is experimentally challenging. Such constrictions are difficult to fabricate, they are very sensitive to electrostatic disharge, and the sample must be brought very close to the constriction (80 nm).

\section{Conclusion}
We have provided the first demonstration of magnetic resonance force microscopy using a membrane resonator. We measured the signal from a 20 $\mu$m DPPH particle placed on a membrane. The membrane exhibited a room temperature sensitivity of 4 fN/$\sqrt{\mathrm{Hz}}$, and a potential low temperature sensitivity of 25 aN/$\sqrt{\mathrm{Hz}}$. Membranes are a practical mechanical resonator for use in MRFM because they are commercially available, low cost, and have high sensitivity. Furthermore they are versatile in that they have a large surface area, high resonant frequency, and can be used in the fields of quantum optomechanics, TEM, and MRFM. Given these properties, membranes are a compelling candidate for a number of applications such as a compact detector that can be used to identify or image nuclear spins in microscopic samples. 

\begin{acknowledgments}
The research presented in this Article was supported by the Army Research Office (grant no. W911NF-09-1-0147), and the Center for Emergent Materials (CEM), an NSF-funded MRSEC through grant DMR-1420451.  Technical support was provided by the NanoSystems Laboratory at The Ohio State University. 
\end{acknowledgments}

\section{Appendix}
To fit the cyclic saturation data of figure \ref{fig:Power_dependence}, we calculate the net force acting on the DPPH particle following reference \cite{Wago1997}.  Frequency modulation causes the resonant slice to oscillate spatially. The resonant slice is 50 $\mu$m in thickness, and the particle is 20 $\mu$m in size. As the magnet approaches the particle, the leading edge of the slice begins to cyclically enter and exit the particle. This creates a force which resonantly drives the membrane. Once the resonant slice is fully inside the particle, there is no net force because spins on either side of the resonant slice create oscillating forces with opposite phase, yielding cancellation of the net force.  As the resonant slice exits the other side of the particle, a force is created 180 degrees out of phase with the drive, which the lock-in records as a negative value.

We model the magnet as a sphere that generates a dipole field $B_0=\frac{\mu_0}{4\pi}
\frac{ m}{\left(x-x_{\mathrm{magnet}}\right)^3}$ and corresponding gradient, $G=\frac{-3\mu_0}{4\pi}
\frac{ m}{\left(x-x_{\mathrm{magnet}}\right)^4}$, where $m=\frac{4}{3}\pi  R^3 M$, $M=1.06\times10^6$ is the magnetization of the NeFeB magnet, and $R\sim 1$ cm is the radius of the magnet. We assume the DPPH particle has roughly a constant cross sectional area $A$. We thus have an equation for the total force on the DPPH particle, which we use to fit the data in figure \ref{fig:Power_dependence},
\begin{multline}
F(x_{\mathrm{magnet}})=\int_{x_{\mathrm{start}}}^{x_{\mathrm{end}}}G\,M_1\, A\, \mathrm{d}x=   \\
\frac{A\,B_1^2\,\gamma\, M_0\,\tau^2\,\Omega}{1+B_1^2\gamma^2\tau^2+\tau^2 \left( \omega_0 - \frac{\mu_0\gamma MR^3}{3\left(x_{\mathrm{magnet}}-x_{\mathrm{start}}\right)^2}\right)^2} \\
-\frac{A\,B_1^2\,\gamma\, M_0\,\tau^2\,\Omega}{1+B_1^2\gamma^2\tau^2+\tau^2 \left( \omega_0 - \frac{\mu_0\gamma MR^3}{3\left(x_{\mathrm{magnet}}-x_{\mathrm{end}}\right)^2}\right)^2},
\label{equation:cyc_sat_fit_eqn}
\end{multline}
where $x_{\mathrm{start}}$ and $x_{\mathrm{end}}$ denote the position of the DPPH particle, and $x_{\mathrm{magnet}}$ is the magnet's position. As the resonant slice passes through the DPPH particle, we measure the cyclic saturation signal shown in figure \ref{fig:Power_dependence}. For the second and third harmonics we replace $M_1$ with $M_2$ and $M_3$ in equation \ref{equation:cyc_sat_fit_eqn} and solve for similar expressions, which we use to fit the data in figure \ref{fig:Second_harmonic}.

\bibliographystyle{naturemag}
\bibliography{Magnetic_resonance_force_detection1}
\end{document}